\documentclass[10pt,conference,compsocconf,letterpaper]{IEEEtran} 

\usepackage{times}
\usepackage[pdftex]{color, graphicx}
\usepackage{subfigure}
\usepackage{comment}
\usepackage{algorithm}
\usepackage{algorithmic}
\usepackage{array}
\usepackage{url}

\newcommand{\fixme}[1]{{\em\bf{[FIXME: #1]}}}
\newcommand{\raj}[1]{{\color{red} *** {\textbf{Rajesh:}}\color{blue}{#1}}{\color{red}***}}
\newcommand{\deepak}[1]{{\color{red} *** {\textbf{Deepak:}}\color{blue}{#1}}{\color{red}***}}

\newcommand{\oq}{``}
\newcommand{\cq}{''}

\title{DAPriv: Decentralized architecture for preserving the privacy of medical data.}

\author
       {Rajesh Sharma$^{1,2}$, Deepak Subramanian$^3$, Satish N. Srirama$^1$
       \\
       $^1$Institute of Computer Science, University of Tartu, Estonia.\\
       \{rajesh, satish.srirama\}@ut.ee\\
       $^2$University of Bologna, Italy.\\
       rajesh.sharma@unibo.it\\
       $^3$CIDRE research group, Sup\'{e}lec, France.\\
       subudeepak@gmail.com\\
       }
       
\begin{document}
\maketitle

\begin{abstract}
The digitization of the medical data has been a sensitive topic. In modern times laws such as the HIPAA provide some guidelines for electronic transactions in medical data to prevent attacks and fraudulent usage of private information. In our paper, we explore an architecture that uses hybrid computing with decentralized key management and show how it is suitable in preventing a special form of re-identification attack that we name as the re-assembly attack. This architecture would be able to use current infrastructure from mobile phones to server certificates and cloud based decentralized storage models in an efficient way to provide a reliable model for communication of medical data. We encompass entities including patients, doctors, insurance agents, emergency contacts, researchers, medical test laboratories and technicians. This is a complete architecture that provides patients with a good level of privacy, secure communication and more direct control.

\end{abstract}





\section{Introduction }
Medical data is one of the most important data sets that have ever been collected with regard to humans. There is a need to share some of this data in lieu of supporting further research. However, in the due process, if the medical records are not being masked properly, then it might result in divulging the private information of the patients. It must be noted that although few patients are known to have been harmed by security breaches of medical computers or devices, the security of medical devices is not a luxury \cite{Maisel2010}. Malicious access for various reasons such as insurance frauds could plague the system. Thus, it is very important that the data stored at these servers must be secured and shared carefully so that the privacy of the patients does not get compromised.

Patients generally trust while interacting with the hospitals. However, researchers can club with hospitals to get medical records of the patients. We present an example to provide the relevant background to the problem. Consider a scenario, where a doctor who is a specialist in a particular medical field and is seeing a large number of patients. Generally, the doctors/hospitals maintain record of the patients. Researchers are generally interested in the data records being maintained by the doctors/hospitals as these researchers can perform data mining operations on these medical records of the patients to get some interesting results or insight which could be useful for devising new tests or drugs. However, in the due process the records of the patients become vulnerable to privacy breaches. We agree that these data mining/research operations are helpful in the field of medical science. However, at the same time the privacy of the patients cannot be ignored. Thus a system where a patient is the sole owner of his/her medical records is the need of time. The incentive for the patient could be monetary or in some cases better facilitation. For example, the collection of such data is more common for new drugs in the market and usually the provider of the drugs would be more willing to see its effects for a price. Thus, we propose an envisioned architecture through which patients remain the owner of their data. In addition to this we also present a mechanism by which researchers can gather medical records of the patients after getting explicit consent of the patients. The mechanism itself would not be transparent to the user and it is completely envisioned to be automated similar to the implementations of banking systems.

In the past researchers have proposed various mechanisms (see section \ref{sec:relwork} for details) to counter various kind of privacy attacks (see section \ref{sec:Intro:prelim} for details) on the medical data. Some of the researchers have proposed new architectures such as \cite{Boyd2006} and, \cite{He2012} while others have proposed various data anonymizing techniques such as \cite{Sweeney2002}, \cite{Machanavajjhala2007} and \cite{Wang2004}, for the medical records before releasing the data for the research purpose. In contrast to this, Gritzalis et. al. proposed new guidelines for the medical field as a privacy solution \cite{Gritzalis2005}.

To protect the privacy of the patients various laws and acts have been formed such as Health Insurance Portability and Accountability Act (HIPAA) in US, Data Protection Act (DPA) in Europe Union and Personally Controlled Electronic Health Records Act 2012 in Australia, to name a few. These laws enforces that physician or the health providers take proper procedures so that the privacy of the patients are protected. In the next section (Preliminaries, see section \ref{sec:Intro:prelim}), we describe various kinds of attacks which are possible to breach the privacy of the data. We next describe a special kind of attack called re-assembly attack, which is a special kind of re-identification attack. The paper proposes an envisioned architecture for the protection of medical records from re-assemble attack.

\subsection{Re-assembly attack}
\begin{figure}[ht]
  \begin{center}
    \subfigure{\label{fig:center}\includegraphics[width=0.8\columnwidth, height=2.2in]{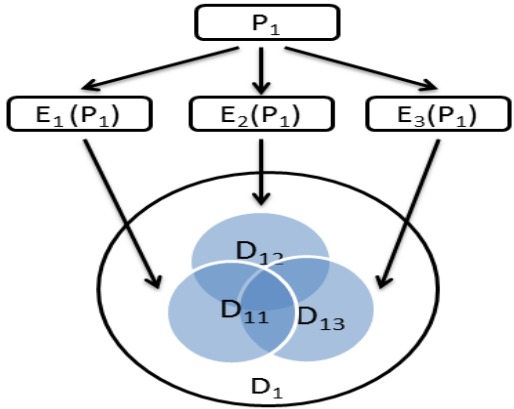}}

  \end{center}
  \caption{Reassemble Attack.}
  \label{fig:reassembleAtk}
\end{figure}

Re-assemble attack occurs in a scenario where various entities collude to create rich information from the chunk of information they contain. These entities try to combine their respective information based on a common identifier such as the Social Security Number (SSN). The information being created can reveal some private information such as location, age or sex about the patient. We describe the re-assemble attack with the help of Figure \ref{fig:reassembleAtk}, which could be a source of privacy leaks. A patient $P_{1}$ might have to interact with various entities like lab personal (let's call it $E_{1}$), nursing personal (let's call it $E_{2}$), etc. During these interactions a patient might have to divulge some personal information (lets call it $D_{11}$, $D_{12}$, etc) during medical tests or day to day routine to these respective entities. It might happen that these entities might collude among themselves to create more information about a patient. The new information $D'_{1}$=$\{$ $D_{11}$, $D_{12}$, $D_{13}$ $\}$ as a result of collusion of various interacting entities might be proper subset of $D_{1}$. However, the chances are that $D'_{1}$ is richer and has more information than the individual information of the interacting entities. The new information $D'_{1}$ might be subject of interest to researchers or attackers. In this paper, we have presented a mechanism to defeat the reassemble attack.


The major contributions of the proposed mechanism in this paper are:
\begin{enumerate}
\item We identify a special kind of re-identification attack, which we term as re-assembly attack.
\item To defeat the re-assembly attack, we present a decentralized architecture which provides full ownership of the medical records to the patients. Hence presents a privacy cover to the patients. We also provide a mechanism by which medical record of the patients can be shared with the researchers anonymously and without affecting the privacy of the patients.
\end{enumerate}

The rest of the paper is organized as follows. Section \ref{sec:Intro:prelim} presents a collection of terms to be used in this paper. In section \ref{sec:relwork}, we present relevant related works. We present our envisioned architecture including various concepts and algorithm in section \ref{sec:approach}. We draw our conclusions and outline our ongoing and planned extensions of DAPriv in Section \ref{sec:concl} .

\section{Preliminaries}\label{sec:Intro:prelim}
In this section we provide brief definitions of various glossary of terms being related and are used in this paper.

The \textbf{least information flow} principle can be defined as, \oq whenever a patient is requested to provide specific personal and/or medical information, she/he runs into the danger of revealing much more information than it is really necessary for the specific task she/he is trying to complete \cite{Gritzalis2005}.



\textbf{Centralized computing} can be defined as computing done at a central location, using terminals that are attached to a central computer \cite{CentralizedWiki}.

\textbf{Decentralized computing} can be defined as the allocation of resources, both hardware and software, to each individual workstation, or office location \cite{DecentralizedWiki, supernova}.

\textbf{Hybrid computing} model works as follows. In a hybrid model, a server is approached first to obtain meta-information, such as the identity of the peer on which some information is stored, or to verify security credentials. From then on, the P2P communication is performed \cite{Milojicic2002}.

\textbf{Interacting entities} are those which share information with each other thereby enhancing each other's knowledge.

\textbf{Quasi-identifiers} refers to sets of attributes (like gender, date of birth, and zip code) that can be linked with external data to uniquely identify individuals in the population \cite{Machanavajjhala2007}.

\textbf{De-identification} or \textbf{Sanitization} is the process by which a collection of data is stripped of information which would allow the identification of the source of the data \cite{DeidentificationWiki, Machanavajjhala2007}.

\textbf{Re-identification attacks} is the process of identifying the source of the data [a.k.a the patient] from documents that have been de-identified \cite{ElEmam2011, Sweeney2002, Bertino2005}.

\textbf{Homogeneity attack} is a specific case of re-identification attack where the patient is deduced due to insufficient diversity in the data pool \cite{Machanavajjhala2007, ElEmam2011}.

\textbf{Background Knowledge attack} is a specific case of re-identification attack where the patient is deduced due to some information about the patient and medical condition provided \cite{Machanavajjhala2007, ElEmam2011}.

\textbf{Linking attack} is a specific case of re-identification attack which refers to a chain of linkages between different available data in various databases to construct a more complete dataset thereby resulting in re-identification \cite{Machanavajjhala2007}. In the linking attack the databases are publicly available. However, there is no publicly available common identifier to link these databases. In contrast to this, in the re-assembly attack, the data is not publicly available. However, the common identifier is publicly available. In other words, various colluding entities try to combine their data with the help of publicly available identifier.

\textbf{Forgery attack} is an attack where one party acts as an authenticated information provider thereby corrupting the database and also in certain cases having successful hacks \cite{Yang2004}.

\textbf{Man-in-the-Middle attack} in cryptography and computer security is a form of active eavesdropping in which the attacker makes independent connections with the victims and relays messages between them, making them believe that they are talking directly to each other over a private connection, when in fact the entire conversation is controlled by the attacker \cite{MITMWiki}.

\section{Literature Review}\label{sec:relwork}
This section describes several interesting research done in the past related to the privacy of the medical records.

\subsection{New guidelines: }
Gritzalis et. al. \cite{Gritzalis2005} present various risks involved with medical data sharing and provide technical guidelines for various protection measures by presenting guidelines that apply privacy-related legislation in a coherent and coordinated way. The work is very useful in understanding the various risks involved and for proper risk management. The paper address the factors of risk with a real-life example with credible supporting data. This is hence important to understand when creating new implementations. However, the scope of the work is limited to providing guidelines to various stakeholders to implement privacy and data protection mechanisms in medical environments.

\subsection{Privacy in medical devices: }Although, in the proposed work we have not taken care of privacy leaks because of medical devices (it is a part of our future work), the medical devices being used by patients can also be a source of privacy leak. Furthermore, these devices poses a new threat since these are susceptible to common types of computer-security breaches such as those caused by computer viruses, internet hackers, and the loss or theft of devices containing sensitive data. Maisel and Kohno \cite{Maisel2010} present a case for the security and privacy of one specific use-case, namely the implantable medical devices. The authors have brought into light the various issues that arise in the field of medical devices and the possibilities of targeted attacks. However, the authors have not provided a model framework to support their arguments. Such a model would have had more weight in completing the paper. On similar lines Maglogiannis et. al. \cite{Maglogiannis2009} also present a mechanism (which they call ``mist'') for one specific use-case namely, to protect location data in patient telemonitoring systems (PTS) from issues pertaining to privacy. The scope of the work is limited to formulating a mechanism to pass location data from PTS while maintaining both privacy and complete functionality to address emergency situations. The work done is very useful in understanding how distributing components helps in privacy. It presents a very convincing case for location protection.

\subsection{Privacy preserving using data anonymization techniques: }
There has also been significant work done in the prevention of re-identification attacks by anonymizing the medical data. Data might be released as long as sufficient metrics for re-identification are not met. K-anonymity is a seminal work by Sweeney \cite{Sweeney2002} in the privacy model for information disclosure which can be used for anonymizing the medical records. An information disclosure provides k-anonymity protection if information for each person contained cannot be distinguished from at-least k-1 individuals. If the database is large enough, and there are enough candidates with shared attributes, then linking becomes more ambiguous. Thus, this model scales better with more data. However, it imposes a huge overload on the system since the system has to maintain the quasi-identifiers. Further, determining k isn't easy though a suitable mechanism has been provided. If k is too low, insufficient ambiguity results and if too high, functionality is affected. Also, K-anonymity is susceptible to homogeneity and background knowledge attacks. Machanavajjhala et. al. \cite{Machanavajjhala2007} present another landmark paper in which they study the concept of l-diversity which presents functionalities not addressed by the k-anonymity model. L-diversity estimates characteristics of data and therefore provides better anonymity. L denotes that entropy of the data [log(l)] satisfies l-diversity. If some specific data from the data pool is deemed insensitive for disclosure, they are not taken into the entropy estimation. However, the model works only if the initial entropy conditions are met. Wang et. al. \cite{Wang2004} propose a methodology for information disclosure using a model of granular computing. The method involves changing the data into a granular form such that each granule contains data that cannot be used to compromise privacy. The model provides a methodology to satisfy the least information flow principle. Each granule is privacy aware and addresses the vital points to privacy. The research calls for some improvements especially in the practicality of the proposed solution. The data anonymization techniques do not take re-assembly attack into account, unlike our proposal. Our architecture ensures that explicit consent of the patient is taken for collecting medical records and for selection of specific data anonymization techniques before data is released.

\subsection{Architecture based approaches: } Some of the researchers have introduced architecture based solutions to tackle the menace of privacy leaks in medical data. The architectures to avoid various attacks can be categorized in following three ways:

\textbf{Centralized model: }In \cite{journals/titb/LeeL08}, authors propose a centralized cryptographic key management framework for Health Insurance Portability and Accountability Act (HIPAA) privacy/security regulations. Patient's data is stored in smart cards. The scope of the work is limited to formulating a centralized key management framework. Only the patient with the valid access control and biometrics can retrieve the key. The authors do not give any reason for having a centralized key management solution. The key is obtained by access control mechanisms, which might make it vulnerable. The work is not conclusive and makes statements based on the HIPAA framework.

\textbf{Hybrid solutions: }Boyd et. al. \cite{Boyd2006} introduces a new entity in their proposed hybrid system which they term as "Honest Broker" (HB) to help in sharing the medical data with research systems. HB would remove sensitive fields such as name before passing them to research systems and refrain from passing research data to systems that contain clinical records. There is a minimal impact on the overall system if individual nodes were compromised. The HB vets data passed to all nodes. However, this presents a "Single point of failure" by centralizing majority of the flow to the honest broker.

\textbf{Decentralized model: }He et. al. \cite{He2012} describe a decentralized trust model in Medical sensor network (MSN) communication. The model can be useful in advancing privacy in the communication. A model of trust recommendations by sensor nodes is framed to prevent the bad-mouthing attack. Trust recommendation is treated separately and only good recommendations are used in the indirect trust model (trust propagation) for communication between two nodes that have not done so before.
A total decentralized mechanism for privacy preserving was presented in \cite{Yang2004}. The authors proposed a smart card enabled privacy preserving e-prescription system. The patient would be able to carry the data using a smart card as a storage mechanism. The device would provide data as required to provide least required information at every given point. All the data is digitally signed to preserve the integrity of the data. However, the paper doesn't address how the data would be communicated to the third parties if the card resides with the user.

\section{DAPriv}\label{sec:approach}
In this section, we enumerate various relevant concepts and describe our proposed architecture for preserving the privacy of medical data.

\subsection{Entities}
In any medical system, there are multiple stakeholders. In this section, we list the various stakeholders in the medical system/environment who interact with patients.

\textbf{Patient: }Any person with a medical record is deemed a patient. In our example, we show an active patient who is seeking medical advise from a physician.

\textbf{Physician: } A physician is an entity who provides information such as medical information and prescriptions to the patient. A physician is bound by the patient-doctor confidentiality and can be trusted with information. A physician could fall under one of the two categories below.

\begin{enumerate}
\item{\textbf{Personal Physician: }Each patient may have a designated personal physician who advises the patient on their medical condition. The personal physician can advice the patient to take up some medical tests, or to consult a specialist physician (termed as remote physician henceforth) or to get admitted in the hospital for specialized treatment. We assumed entities such as nursing staff merely represents the doctors and are bound by the same laws.}

\item{\textbf{Remote Physicians: }These physicians are consulting physicians or specialists. These also include the physicians a person might consult during its travel. A remote physician is the one who is not the permanent personal/family physician of the patient. This class of physicians may not be provided with all of the data. For example, an ENT specialist may not be provided with the data regarding the patients' psychology.}
\end{enumerate}


\textbf{Test Laboratories: }Patients might have to take some medical tests. The lab might have to report the test results to the patients or in some cases to the doctors directly. The results will then be interpreted by the doctors.


\textbf{Insurance Agent: }Users/ patients are generally being covered by insurance. Most of the time these insurance companies require some parts if the medical data to process the payment. In the model, we have not considered this entity as we assumed because of governmental laws insurance agents will not disclose patients data with researchers \cite{HIPAA}.

\textbf{Researchers: }Researchers are the entities who try to do various research operations (for example data mining) to get some insights from medical records. The term researchers could represents an organization or a person. However, we will use the term interchangeably in context of an organization and a person. The researchers may send out invitations only to doctors who in turn may choose to forward such invitations. The researchers are listed in dictionaries and can also be found directly by the patients through advertisements etc.

\subsection{Architecture}
In the introduction, we described how the different entities could collude and perform a re-assembly attack. The attack can expose the sensitive information about the patients. This forms the primary reason for proposing our architecture which is a patient oriented architecture. The onus of data disclosure is on the patient. There can be no release of information about the patient without the explicit consent. Further, we include another step of anonymization to include the works of other researchers in privacy such as k-anonymity \cite{Sweeney2002} and l-diversity \cite{Machanavajjhala2007}.

\subsubsection{Certifying Authority }Certifying authority is an authority which provides public and private keys to various entities in the system. We assume each entity (for example patients, physicians, and lab personnel) have a public and as well as private key. Certifying authority issues these keys to various entities. These keys are integral part of the system which makes sure that the system is privacy proof. They are used typically as signatures and to verify the validity of the source/destination.

\subsubsection{Anonymizers/Sanitizers }This is the entity which enforces the process of anonymizing or sanitizing the data. The patient themselves pre-anonymize the data by removing fields they do not wish to share and sharing only data they wish to. This entity further sanitizes data based on the submission pool. This is done so that the patient may have a better possibility of privacy. The anonymizers are also de-centralized and thus help in enforcing the model. The patient also has the option of not trusting any of the anonymizers and directly submitting all the data. Similarly, the patient can engage the services of an authorization server to have a more anonymous communication with the anonymizer as well.

\subsubsection{Directory Server }These servers act as databases which lists all the laboratories and researchers. When a patient approaches a laboratory for the tests, the lab provides its identification of the form DirectoryID\#LabID. The patient sends this value to the authorization sever which verifies laboratorie(s) through Directory server.


\subsubsection{Authorization Server }Authorization server maintain lists of directory servers. Each directory server is stored with a unique ID. Further, these servers act as a bridge between entities wherever needed. Unlike a tunnel based bridge such as tor \cite{TORWiki}, the authorization server also helps to perform certain key operations. However, as a possibility, the patient may themselves implement the authorization server mechanism and use tor-like tunnels for anonymity of communication.

\subsubsection{Emergency contact }The contact which will access the patient's data in case of emergency for example when the patient is in deep comma or is on run.

\subsubsection{Emergency Contact Storage Server }This server stores the data of the patients however this can only be accessed by patient's emergency contact. The emergency contact server contains the public key of the emergency contact. All the patient data (this does not include data about the various patient keys) are then encrypted and added to the server. The emergency contact may never access the data on the server without triggering a signal to the patient. The signal is triggered in the form of a standard flag and log mechanism.

\subsubsection{Data tampering by the patients }A patient oriented architecture, with all data controlled by the patient, reduces the security and reliability of data for other organizations such as insurance agents. To avoid tampering of the records by patients themselves, any important entity providing data of significance must always digitally sign the data. For example, the digital signature can be used by the doctors to validate the prescription given to a patient.

\subsubsection{Encrypting the patient's data }Data needs to be provided back to the patient. The public key is provided to the interacting entity. For example, in the instance of lab test results, the lab personnel write the data that has been encrypted with the patient's public key, back to the temporary communication storage location. The private keys of the patient are not shared with any other entity / person. They remain private. The patient has a set of private keys and each private key has a few public keys. At every use of a key, a decay rate for the public and private keys are calculated. This is because of two reasons.
\begin{enumerate}
\item{Using the same public key would identify the patient to be unique}
\item{Using multiple public keys for the same private key increases the probability of detection of the private key}
\end{enumerate}
If the threshold is reached for a public key, the respective public key is archived and another public key for the private key is created. If the threshold is reached for a private key, all its public keys are archived and a new private key with a set of public keys is created. To increase complexity and randomness, the patient has multiple private keys, each with multiple public keys. This framework of private keys with multiple subkeys can be accomplish with most common algorithms such as RSA \cite{RSA} or Elliptic Curve \cite{EllipticCurve}. At any given request, a public key is selected at random from all the available sets. Similarly when the patient passes the data to the physician, they are encrypted with the physician's public key.

The mechanism itself however is completely automated and the patient does not have to worry about the complexities of the process analogous to general implementations of banking systems.

\begin{figure*}[ht]
  \begin{center}
    \subfigure{\label{fig:center}\includegraphics[width=0.8\textwidth]{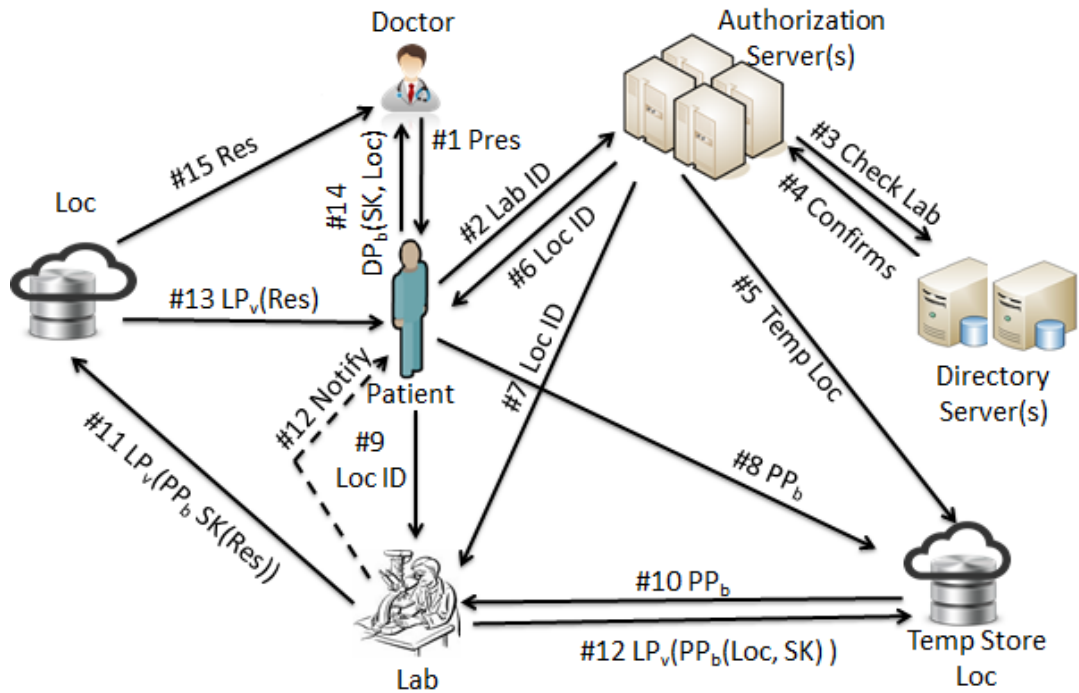}}
  \end{center}
  \caption{Patient Lab Interaction.}
  \label{fig:LabPatient}
\end{figure*}

\subsubsection{How to avoid Re-assembly attack: }
A patient divulges a certain set of information at every lab test and at every point where he/she communicates with a lab personal. Various interacting entities can collude with researchers to get more information about patients. An identifier such as SSN can be used to locate various records of a patient scattered across with various interacting entities. This is named as re-assembly attack, which we classify as a special kind of re-identification attack.


To tackle the re-assembly attack we propose the following architecture (see Figure \ref{fig:LabPatient}). The primary solution is our key management mechanism. A patient can use multiple keys while interacting with various entities. Even if various entities collude they will not be able to identify which record belongs to which patient. With the help of an architectural diagram (see Figure \ref{fig:LabPatient}) we next explain the process to defeat the re-assembly attack in detail.



The edges represent the relation between two entities. Furthermore, in following listing with the help of edge's labels (\#Edge Number) we explain the process how the architecture can be used for the deterrent of re-assembly attack.

\begin{itemize}
\item  \#1: The doctor passes the prescription (Pres), signed by his digital signature to the patient.
\item  \#2: Patient selects a lab for undergoing the tests. The Patient asks the AuthServer to authorize the tests prescribed by the doctor at a lab using the Lab ID.
\item  \#3: The AuthServer contacts Directory Servers to verify if the lab can perform these tests.
\item  \#4: Directorty Servers confirms about the Lab's testing capabilities (verification) back to AuthServer.
\item  \#5: The temp communication store location is created by AuthSever. 
\item  \#6: The location is given to the patient with an identifier.
\item  \#7: In parallel, the location is given to the lab with the same identifier.
\item  \#8: Patient stores its Public key at temp store location.
\item  \#9: Patient approaches the lab with the location identifier and undergoes test(s).
\item  \#10: Lab gets the patient's public key ($PP_{b}$) from the temp store location. The usage of the patient's public key is describe next.
\item \#11: Lab stores the result (Res) of the Patient's medical test (encrypted with a symmetric key (SK) at new location (Loc). This information is further encrypted with the public key of the patient ($PP_{b}$). For verification, lab signed the result file with its digital signature ($LP_{v}$).
\item \#12: Lab stores the information about the result and new location at temp store location. At temp store location, the lab stores the (i) location (loc) where the actual result file is stored and (ii) the symmetric key (SK) to open the file. This information is stored using the public key of the patient ($PP_{b}$). The lab uses its digital signature ($LP_{v}$) to store this information. Lab also notifies the patient about the placement of the result at new location.
\item \#13: Once the patient is notified about the test results, it goes to the temp store location to fetch the location where the actual test file is stored. Patient uses his/her private key to decrypt the information.
\item \#14: Patient can then deliver then notifies the doctor (D) about the result. It passes the location (loc) where the actual result is stored and as well as the symmetric key to open the file. This information is passed to doctor using the public key of the doctor ($DP_{b}$).
\item \#15: Doctor fetches the result file from the information provided to it by the patient.
\end{itemize}


\begin{figure*}[ht]
  \begin{center}
    \subfigure{\label{fig:center}\includegraphics[width=0.8\textwidth]{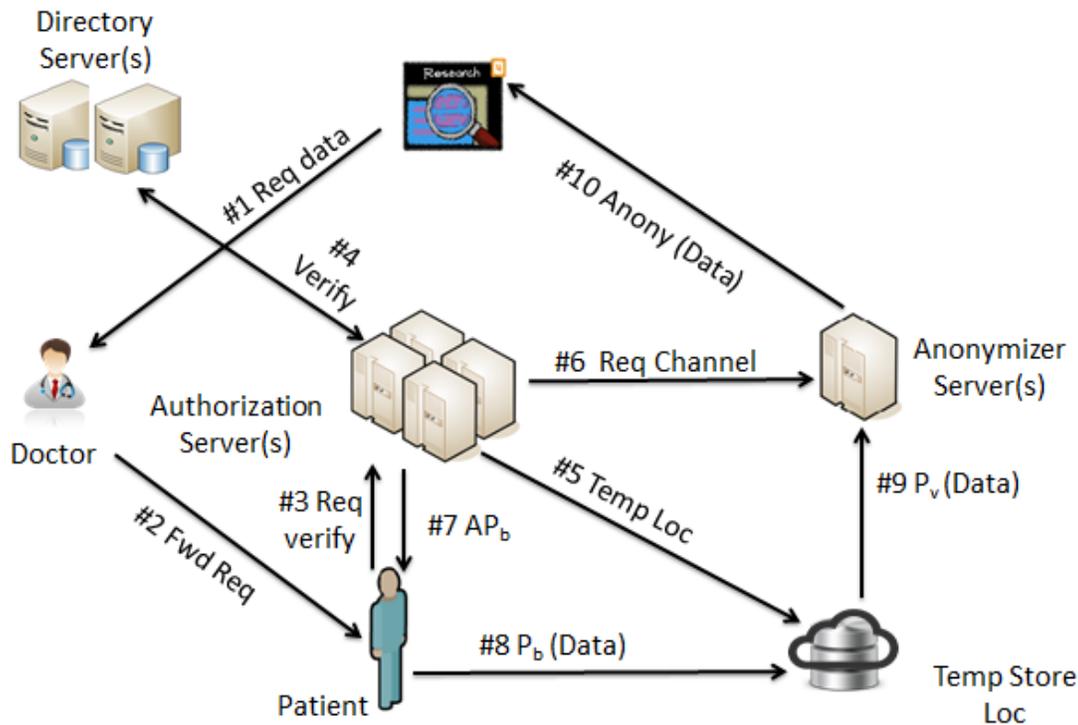}}
  \end{center}
  \caption{Gathering of data by Researcher.}
  \label{fig:ResearchData}
\end{figure*}

\subsubsection{Gathering of medical records for research: }
Our proposed architecture provides a mechanism for the privacy leaks however, we do understand for the medical research often medical records in turn can benefit the patients themselves. In Figure \ref{fig:ResearchData}, we present the process by which researchers can requests patients through their personal doctors. With the help of figure \ref{fig:ResearchData} and following steps we describe the process by which medical data can be gathered by researchers.


\begin{itemize}
\item \#1: Researcher requests the doctors for the medical records.
\item \#2: Doctor/Physician forwards the request to the patients with the researcher's identifier.
\item \#3: Patient contacts the AuthServer for verification of the researcher.
\item \#4: AuthServer further contact the directory server(s) for verification of the researcher.
\item \#5: After verification, AuthServer creates a temp store location.
\item \#6: Also, AuthServer requests Anonymizer for creating a anonymizer channel.
\item \#7: AuthServer updates the patient with the public key of the Anonymizer.
\item \#8: Patient sanitizes the data, that is, removes the fields which it does not want to divulge. It then encrypt the data using the public key of the anonymizer.
\item \#9: Anonymizer decrypts the data and apply data anonymizing techniques on the set of data it has collected.
\item \#10: Anonymizer then forwards the data to researcher.
\end{itemize}

In a decentralized architecture it is very difficult to compromise all the entities (Authorization server, doctors etc) at the same time as compared to centralized architecture \cite{Boyd2006}. Thus, by distributing the responsibilities and assigning different roles to different entities, the architecture helps in the protection of the privacy of the patients.

\section{Conclusion and Future Work}\label{sec:concl}
Medical records of the patients are a valuable asset for performing the research. Equally important is the privacy of the patients whose medical records are investigated for the research. Even if the patient's data is secure with its personal doctors, the medical data of the patient can be recovered through various entities dealing with patients. We term this attack as re-assembly attack where various entities can collude to get more information about the medical history/data of the patients. In the due process this can have an impact on the privacy of the patients. We present a mechanism by which the re-assembly attack can be avoided. We argue that to protect the privacy of the medical records the patients should be given full ownership and control of their data. Patients can decide themselves what fields to expose to the researchers. To achieve this goal, we presented a privacy preserving, patient oriented architecture. The architecture forces the researchers to take consent of the patients (through their personal doctors) before they can obtain the desired data.

The privacy of the medical records can be achieved with the combined efforts of proper laws and technical solutions. The proposed architecture takes care of the present medical laws and procedures in various countries and thus the architecture assumes personal doctors  and insurance agents will not share the medical records of their patients. However, we accept that more robust improvements are required to defeat the privacy problems of the medical records. One approach could be to provide the sole ownership of the medical records to the patients themselves. To achieve this we need a complete decentralized architecture on the lines of \cite{supernova} where patients store their data on their systems. This is analogous to the concept of Bring Your Own Device (BYOD) as proposed in \cite{Yang2004}. The patients can carry the smart device themselves and can give temporal access to the interacting medical entities. However, this idea might not be feasible for various reasons. For example, non-privileged patients or for situations like emergency. The solution for these scenarios could be replication of data from the patient's device and the introduction of emergency contacts.

Nonetheless, our architecture can further be improved in following ways:
\begin{enumerate}
\item In the present architecture a patient can send someone else to the labs to undergo tests. To avoid impersonation by the patient a proper check should be in place.
\item Also the architecture needs to extend to cover insurance agents as well.
\item Currently we have not put alarms in case an entity is compromised. Proper secure alarms are need to be put so that entities can be updated in case a particular entity gets compromised.
\item An extension of the present work also plans to cover the solution for the privacy breach because of the medical devices.
\end{enumerate}

\section{Acknowledgments}
The research is supported by the Estonian Science Foundation grants ETF9287 and PUT360, and the European Social Fund for Doctoral Studies and Internationalisation Programme DoRa.

\bibliographystyle{abbrv}
\bibliography{disco}

\begin{thebibliography}{10}

\bibitem{Bertino2005}
E.~Bertino, B.~C. Ooi, Y.~Yang, and R.~H. Deng.
\newblock {Privacy and ownership preserving of outsourced medical data}.
\newblock In {\em Proceedings of the 21st International Conference on Data
  Engineering, 2005. ICDE 2005.}, number Icde, pages 521-- 532. IEEE Xplore,
  2005.

\bibitem{Boyd2006}
A.~D. Boyd, C.~Hosner, D.~a. Hunscher, B.~D. Athey, D.~J. Clauw, and L.~a.
  Green.
\newblock {An 'Honest Broker' mechanism to maintain privacy for patient care
  and academic medical research.}
\newblock {\em International journal of medical informatics}, 76(5-6):407--11,
  2006.

\bibitem{CentralizedWiki}
C.~computing.
\newblock {http://en.wikipedia.org/wiki/Centralized\_computing}.

\bibitem{DecentralizedWiki}
D.~Computing.
\newblock {http://en.wikipedia.org/wiki/Decentralized\_computing}.

\bibitem{DeidentificationWiki}
De-identification.
\newblock {http://en.wikipedia.org/wiki/De-identification}.

\bibitem{ElEmam2011}
K.~{El Emam}, E.~Jonker, L.~Arbuckle, and B.~Malin.
\newblock {A systematic review of re-identification attacks on health data.}
\newblock {\em PloS one}, 6(12):1--12, Jan. 2011.

\bibitem{Gritzalis2005}
S.~Gritzalis, C.~Lambrinoudakis, D.~Lekkas, and S.~Deftereos.
\newblock {Technical guidelines for enhancing privacy and data protection in
  modern electronic medical environments}.
\newblock {\em Information Technology in Biomedicine, IEEE Transactions on},
  9(3):413--423, 2005.

\bibitem{EllipticCurve}
D.~Hankerson, A.~J. Menezes, and S.~Vanstone.
\newblock {\em Guide to Elliptic Curve Cryptography}.
\newblock Springer-Verlag New York, Inc., Secaucus, NJ, USA, 2003.

\bibitem{He2012}
D.~He, C.~Chen, S.~Chan, J.~Bu, and A.~V. Vasilakos.
\newblock {ReTrust: Attack-Resistant and Lightweight Trust Management for
  Medical Sensor Networks}.
\newblock {\em Information Technology in Biomedicine, IEEE Transactions on},
  16(4):623--632, 2012.

\bibitem{HIPAA}
s.~S.~S. HIPAA~Act, 104th~Congress.
\newblock
  {http://www.gpo.gov/fdsys/pkg/CRPT-104hrpt736/pdf/CRPT-104hrpt736.pdf}.

\bibitem{journals/titb/LeeL08}
W.-B. Lee and C.-D. Lee.
\newblock A cryptographic key management solution for hipaa privacy/security
  regulations.
\newblock {\em IEEE Transactions on Information Technology in Biomedicine},
  12(1):34--41, 2008.

\bibitem{Machanavajjhala2007}
A.~Machanavajjhala, D.~Kifer, J.~Gehrke, and M.~Venkitasubramaniam.
\newblock {L-diversity: Privacy beyond k-anonymity}.
\newblock {\em ACM Transactions on Knowledge Discovery from Data}, 1(1), mar
  2007.

\bibitem{Maglogiannis2009}
I.~Maglogiannis, L.~Kazatzopoulos, K.~Delakouridis, and S.~Hadjiefthymiades.
\newblock {Enabling Location Privacy and Medical Data Encryption in Patient
  Telemonitoring Systems}.
\newblock {\em Information Technology in Biomedicine, IEEE Transactions on},
  13(6):946--954, 2009.

\bibitem{Maisel2010}
W.~Maisel and T.~Kohno.
\newblock {Improving the security and privacy of implantable medical devices}.
\newblock {\em New England journal of medicine}, 362(13):1164--1166, 2010.

\bibitem{Milojicic2002}
D.~S. Milojicic, V.~Kalogeraki, R.~Lukose, K.~Nagaraja, J.~Pruyne, B.~Richard,
  S.~Rollins, Z.~Xu, D.~S. Milojicic, V.~Kalogeraki, R.~Lukose, K.~Nagaraja,
  J.~I.~M. Pruyne, B.~Richard, S.~Rollins, and Z.~Xu.
\newblock {Peer-to-Peer Computing}.
\newblock Technical report, HPL, Hewlett-Packard, 2002.

\bibitem{MITMWiki}
MITM.
\newblock http://en.wikipedia.org/wiki/man-in-the-middle\_attack.

\bibitem{RSA}
R.~Rivest, A.~Shamir, and L.~Adleman.
\newblock A method for obtaining digital signatures and public-key
  cryptosystems.
\newblock {\em Communications of the ACM}, 21:120--126, 1978.

\bibitem{supernova}
R.~Sharma and A.~Datta.
\newblock Supernova: Super-peers based architecture for decentralized online
  social networks.
\newblock In {\em COMSNETS}, 2012.

\bibitem{Sweeney2002}
L.~Sweeney.
\newblock k-anonymity: a model for protecting privacy.
\newblock {\em International Journal of Uncertainty, Fuzziness and
  Knowledge-Based Systems}, 10(5):557 -- 570, 2002.

\bibitem{TORWiki}
{TOR}.
\newblock http://en.wikipedia.org/wiki/tor\_(anonymity\_network).

\bibitem{Wang2004}
D.-W. Wang, C.-J. Liau, and T.-S. Hsu.
\newblock {Medical privacy protection based on granular computing.}
\newblock {\em Artificial intelligence in medicine}, 32(2):137--49, oct 2004.

\bibitem{Yang2004}
Y.~Yang, X.~Han, F.~Bao, and R.-H. Deng.
\newblock {A smart-card-enabled privacy preserving E-prescription system}.
\newblock {\em Information Technology in Biomedicine, IEEE Transactions on},
  8(1):47--58, 2004.

\end{thebibliography}
\label{sec:References}
\end{document}